\documentclass[showpacs,fleqn,nobibnotes,preprint]{revtex4}
\usepackage{amsmath}
\usepackage{graphicx}
\usepackage{float}
\usepackage[utf8]{inputenc}

\usepackage{subfigure}

\usepackage{multirow}

\usepackage{soul}

\def\lsim{\raise0.3ex\hbox{$<$\kern-0.75em\raise-1.1ex\hbox{$\sim$}}}
\def\gsim{\raise0.3ex\hbox{$>$\kern-0.75em\raise-1.1ex\hbox{$\sim$}}}

\def\pom{{I\!\!P}}

\newcommand{\be}{\begin{equation}}
\newcommand{\ee}{\end{equation}}

\begin{document}

\title{Investigating the influence of diffractive interactions on ultra - high energy extensive air showers}
%~ \pacs{12.38-t; 13.85.-t; 13.15.+g}
\author{L.B. Arbeletche, V.P. Gon\c{c}alves and M.A. M\"uller}
\affiliation{High and Medium Energy Group \\ Instituto de F\'{\i}sica e Matem\'atica, Universidade Federal de
Pelotas\\
Caixa Postal 354, CEP 96010-900, Pelotas, RS, Brazil.
}

% ==================================================================== %
\begin{abstract}

The understanding of the basic properties of the ultra - high  energy extensive air showers is strongly dependent on the description of the hadronic interactions in a energy  range beyond that probed by the LHC. One of the uncertainties present in the modeling of the air showers is the treatment of diffractive interactions, which are dominated by non - perturbative physics and usually described by phenomenological models. These  interactions are expect to affect the development of the air showers, since they provide  a way of transporting substantial amounts of energy deep in the atmosphere, modifying  the global characteristics of the shower profile. In this paper we investigate  the impact of the diffractive interactions in the observables that can be measured in hadronic collisions at high energies and ultra - high energy cosmic ray interactions. We consider three distinct phenomenological models for the treatment of  diffractive physics and estimate the influence of these interactions on the elasticity, number of secondaries, longitudinal air shower profiles and muon densities for proton - air and iron - air collisions at different primary energies. Our results demonstrate that the diffractive events has a non - negligible effect on the observables and that the distinct approaches for these interactions, present in the phenomenological models, are an important source of theoretical uncertainty for the description of the extensive air showers.   

	%~ The high parton density present at  high energies and large nuclei is expected to modify the lepton - hadron cross section and the associated observables. In this paper we analyse the impact of the high density effects in the average inelasticity and the neutrino - nucleus cross section at ultra high energies. We compare the predictions associated to the linear DGLAP dynamics with those from the  Color Glass Condensate formalism, which includes non-linear effects. Our results demonstrated that the non-linear effects reduce the average inelasticity and that the predictions of the distinct approaches for the neutrino - nucleus cross section at ultra-high energies are similar.

\end{abstract}
% ==================================================================== %

\maketitle

% ==================================================================== %
\section{Introduction}
\label{intro}

Understanding the behaviour of high energy hadron reactions from a fundamental perspective within Quantum Chromodynamics (QCD) is an important goal of particle physics. One of the main open questions is the treatment of the diffractive processes, which are characterized by the presence of an intact hadron and large rapidity gaps in the final state (For a recent review see Ref. \cite{forward}). These processes are in general described in terms of a color singlet object: the Pomeron ($\pom$). This object, with the vacuum quantum numbers, was introduced phenomenologically in the Regge theory as a simple moving pole in the complex
angular momentum plane, to describe the high-energy behaviour of the total and elastic cross-sections of the hadronic reactions \cite{collins}. 
The diffractive events are dominated by low transverse momentum processes, i.e. processes in which the strong running coupling constant is large and the useful perturbative methods are not valid. These processes are in general described by phenomenological models based on first principles of the Quantum Field Theory and basic properties of QCD \cite{kaidalov}. It implies a large theoretical uncertainty, with a strong impact on the predictions for the magnitude and energy dependence of the diffractive cross section. For example, while some models \cite{qgsjet} predict that $\sigma_{diff} \propto \ln^2 s$ at asymptotic energies, other models \cite{sibyll} predict  $\sigma_{diff} \propto \ln \, s$. As a consequence, the contribution of the diffractive events for the total cross sections at very high energies still is an open question. Recent experimental results from the Run 1 of the LHC have shed some light on the energy behaviour of the single and double diffractive cross sections \cite{lhcdata} and more precise results are expected in the Run 2 \cite{forward}. The expectation is that these data could be used to constrain the basic assumptions present in the phenomenological models, decreasing the uncertainty in its predictions at larger energies.

The understanding of the hadronic interactions also is fundamental for the description of the ultra - high energy cosmic ray (UHECR) air showers \cite{annual,lipari}, with the reconstruction of the primary UHECR properties being  strongly  dependent  on the treatment of the diffractive and non-diffractive events present in the hadron -- air interactions. Due to the dominance of soft physics, the models of hadronic interactions in the generator models used in the simulation of extensive air showers (EAS) are largely phenomenological and have to be extrapolated from accelerator energies, where they are calibrated, to the UHECR energies. In recent years, the underlying theoretical framework present in these generators have been improved and  the LHC data have been used to tuning  the basic cross sections \cite{qgsjet,sibyll23,epos}. However, the available collider data do not cover the full kinematic region of interest in UHECR interactions \cite{ralph_epj}. In particular, experimental data for the particle production at very forward rapidities still are scarce, with its theoretical description still being an open question (For a recent study see, e.g. Ref. \cite{werner}). As a consequence of the theoretical uncertainty present in the description of the hadronic interactions present in the EAS, the air  shower simulations still are one of the main source of systematic uncertainty in the interpretation of cosmic  ray data \cite{david,parsons,auger}.

Our main goal in this paper is to give a quantitative estimate of the uncertainty associated to the treatment of the diffractive interactions on the shower observables. 
These interactions are expect to affect the development of the air showers, since  diffractive interactions have direct impact on the inelasticity -- the relative energy loss of the leading secondary particle --  providing  a way of transporting substantial amounts of energy deep in the atmosphere. Consequently, they modify the global characteristics of the shower profile. In particular, a higher diffraction rate implies a slower EAS development, modifying the position of the shower maximum $\langle X_{max}\rangle$, and a smaller number of secondaries in each interaction. 
In what follows we will study the impact of the diffractive interactions on ultra - high energy extensive air showers through the comparative analysis of the predictions for the EAS observables from the standard versions of the Sibyll \cite{sibyll}, QGSJET \cite{qgsjet} and EPOS \cite{epos} hadronic models, which are available in the framework of the CORSIKA air shower simulation package \cite{corsika}. Our study is strongly motivated by the analysis performed thirteen years ago in Ref. \cite{garcia1}, where the authors have studied the same problem using the pre - LHC hadronic models and the  AIRES program \cite{aires} for the simulation of the EAS development. In what follows we will update that analysis by considering the current hadronic models  and include new results for  Fe-Air collisions which were not considered in that reference.   In order to investigate the contribution of these interactions,    we will present, for each hadronic model,  a comparison between predictions obtained using the full simulations, i.e. including the non - diffractive and diffractive events, with those derived excluding the diffractive events of the EAS development. In particular, in Section \ref{hadron} we will compare the predictions of the different hadronic models for the distribution of number of secondaries, fraction of diffractive events, average fraction of pions and elasticity distributions, considering  individual $p$-Air and $Fe$-Air collisions at different values for the primary energy. The impact on the EAS observables is studied in Section \ref{shower}, where 
we will analyse the impact of the diffractive interactions on the longitudinal profiles of charged particles and muons as well as on the position of the shower maximum. 
Finally, in Section \ref{conc} we summarize our main conclusions.

% ==================================================================== %
\section{Impact of the diffractive interactions in hadronic collisions}
\label{hadron}

As discussed in the Introduction, the treatment of the hadronic interactions in the air shower simulation codes is based on phenomenological models. In what follows we will consider the Sibyll 2.1, QGSJET -- II 04 and EPOS LHC models, present in the CORSIKA package, and compare its predictions for distinct observables in  $p$-Air and $Fe$-Air collisions at different center-of-mass energies. Before to present our results, lets present a brief review of the main assumptions of the different phenomenological models (For a more detailed review see, e.g., Ref. \cite{david}). 

Most current hadronic interaction models are based on basic quantum  field theory principles, such as unitarity and analyticity of scattering amplitudes, and use Gribov - Regge theory \cite{gribov} of multi - Pomeron exchange between nucleons as the basis for the treatment of high energy, soft interactions. Perturbative QCD is considered to describe hard interactions with high transverse momentum, which becomes important at high energies. In general, simplifications are made in the implementation of the hard processes if they are not directly relevant to the production of high energy secondaries. QGSJET and Sibyll consider the eikonal model and assume  unitarized cross sections with the real eikonal function being given by a sum of soft and hard contributions. EPOS also is based on Gribov - Regge theory and provides a energy - conserving quantum mechanical multiple scattering approach in terms of parton ladders.
At high energies, nonlinear effects associated to the high partonic density becomes important and should be taken into account \cite{glr,hdqcd}.  QGSJET provides a microscopic treatment of nonlinear interaction effects in hadronic and nuclear collisions in terms of Pomeron - Pomeron interaction diagrams. On the other hand, in Sibyll these effects are modelled by means of an energy - dependent cutoff for the minijet production. In contrast, 
EPOS employs a phenomenological approach for nonlinear interaction effects and address the energy - momentum correlations between multiple scattering processes at the amplitude level. The generalization from $pp$ to $pA$ and $AA$ collisions is usually performed via Glauber - Gribov formalism \cite{glauber,gribov2}, taking into account inelastic screening and low mass diffraction effects. Finally, diffractive interactions are treated differently in the distinct models. In Sibyll and QGSJET models, the diffraction dissociation is described in terms of the Good - Walker formalism \cite{gw}, where the colliding hadrons are represented by superpositions of elastic scattering eigenstates which undergo different absorption during the collision. In Sibyll, high mass diffraction is described in terms of a 2 - channel eikonal approach. On the other hand, the description of this process in QGSJET is based on all - order resummation of cut enhanced $\pom$ - diagrams. In contrast,  a particular kind of Pomeron is used to define a diffractive event in EPOS. Depending of each event configuration it can be classified as non - diffractive, low mass diffraction without central particle production, or high mass diffraction.

	A basic feature of the diffractive interactions  is that there is a leading particle whose energy is much larger than the energies of the other particles  and the total number of secondaries is generally small. The resulting final states have a high elasticity. In previous studies, these characteristics were used to tag the diffractive events \cite{garcia1,garcia2}. In contrast, in our study the classification of the diffractive and non - diffractive events was made using internal variables of the respective hadronic generators. In particular, in the EPOS LHC and QGSJET-II  04 generators, the events were classified according to the variable {\fontfamily{pcr}\selectfont typevt}, which define the collision type: if {\fontfamily{pcr}\selectfont typevt = 1}, the event is classified as non-diffractive; otherwise, the event is considered a diffractive one. As this variable is read after the event is generated, it allows to reject  diffractive events when generating samples of non-diffractive events.
In contrast, in Sibyll 2.1,  a hadron-air interaction is classified as diffractive requiring that  there is only one wounded nucleon in the target ({\fontfamily{pcr}\selectfont NW = 1}) and that the interaction with this nucleon is diffractive ({\fontfamily{pcr}\selectfont JDIFF(0) > 0}). Again, variables are read after the event have been generated. For nucleus-nucleus interactions, we implemented a distinct scheme for this model. As Sibyll 2.1 treats nucleus-nucleus collisions in a semisuperposition model, where the \textit{A} projectile nucleons are considered independent particles, it generates \textit{A} superposed interactions. We found reasonable to classify as diffractive the collisions whose all superposed interactions were diffractive. Otherwise, if at least one interaction is  non-diffractive, the event is labelled as non-diffractive.

\begin{figure}[t]
		\includegraphics[width=5 cm]{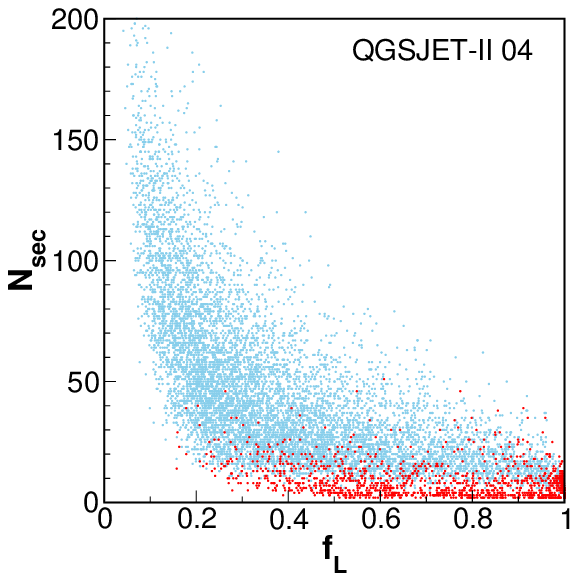}
		\includegraphics[width=5 cm]{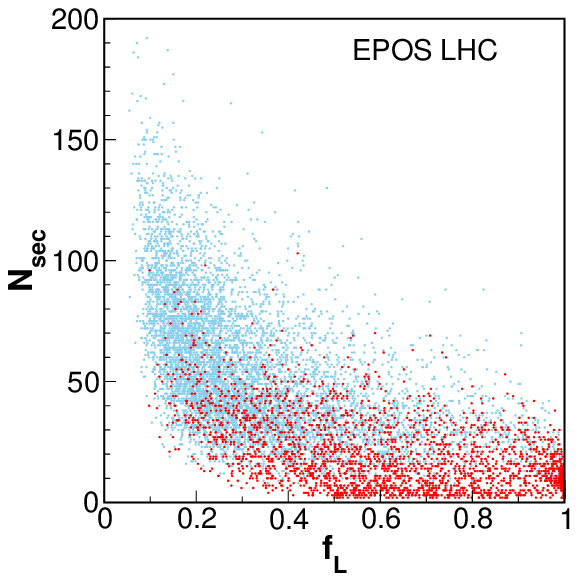}
		\includegraphics[width=5 cm]{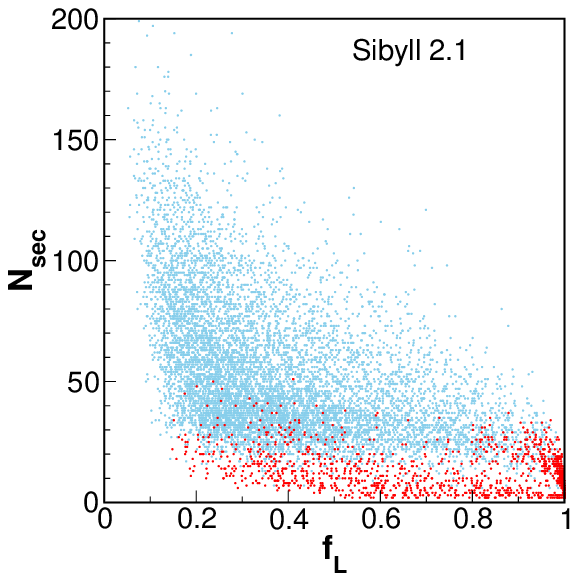}
		\caption{Predictions of the different hadronic generators for diffractive and non - diffractive events in the elasticity -- multiplicity ($f_L \times N_{sec}$) plane considering $10^{4}$ $p$ - Air collisions at $\sqrt{s} = 10$ TeV. Diffractive events are represented by red squares and the non-diffractive one by  light blue circles.}
		\label{tagging}
	\end{figure}

	In order to illustrate the classification  between diffractive and non - diffractive events, in Fig. \ref{tagging} we present our results for the elasticity-multiplicity plane considering $10^{4}$ $p$ - Air collisions at $\sqrt{s} = 10$ TeV. The elasticity is characterized by the leading energy fraction $f_L$, which is defined by the ratio between the energy of the secondary with maximum energy $E_{lead}$ (leading particle) and the primary energy $E_{prim}$, i. e. $f_L \equiv E_{lead}/E_{prim}$.  The predictions of the three different hadronic generators are presented separately, with the  diffractive events being represented by red squares and the non-diffractive one by  light blue circles. We have that the  most of the diffractive events appear on the region of low multiplicity and elasticity close to 1, in agreement with the theoretical expectation. However, the distribution is different for the distinct hadronic models. In what follows, we will analyse the phenomenological implications of these differences.

	We have generated samples of $10^{4}$ $p$ - Air and $Fe$ - Air collisions at different center of mass energies for each event generator model: QGSJET-II 04, EPOS LHC and Sibyll 2.1. 
We assume that the  target is a mixture of nitrogen, oxygen and argon, exactly as it is implemented in the CORSIKA package \cite{corsika}. After the events have been generated, we register all secondaries, imposing a cutoff on the kinetic energy of the secondary particle. Basically,  we excluded from the secondary list all particles with kinetic energy smaller than 40 MeV. It is assumed to eliminate from the secondary list, the fragments of the target nucleus, which are included in the EPOS LHC case, but not in the others models. Such cutoff is expected to have  no influence in the air shower development \cite{garcia2}, since  particles with such low values of kinetic energy should not be able to propagate in the atmosphere. 	{ As} we are interested in the particular case of diffractive collisions, we generated separately samples without diffractive events and full samples including non - diffractive and diffractive events. In what follows we will compare the predictions from these two configurations, which  allow us to estimate the impact of diffraction on the average properties of the hadronic collisions as well as to investigate the theoretical uncertainty associated to the distinct treatment of diffraction present in the hadronic generators considered in our study.
		
		\begin{figure}[t]
		\includegraphics[width=\textwidth]{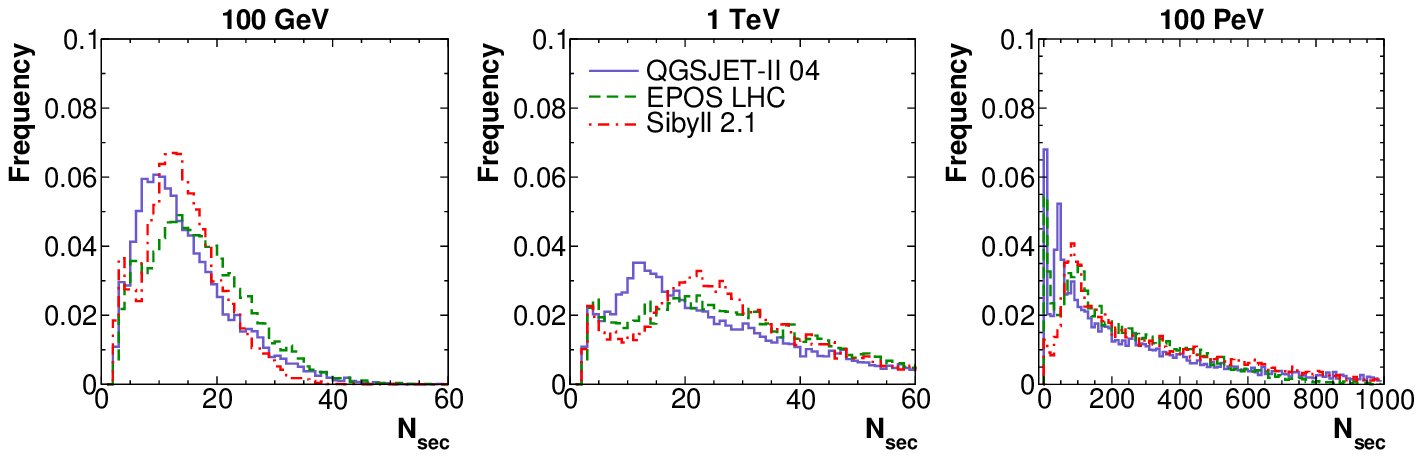} \\
		\includegraphics[width=\textwidth]{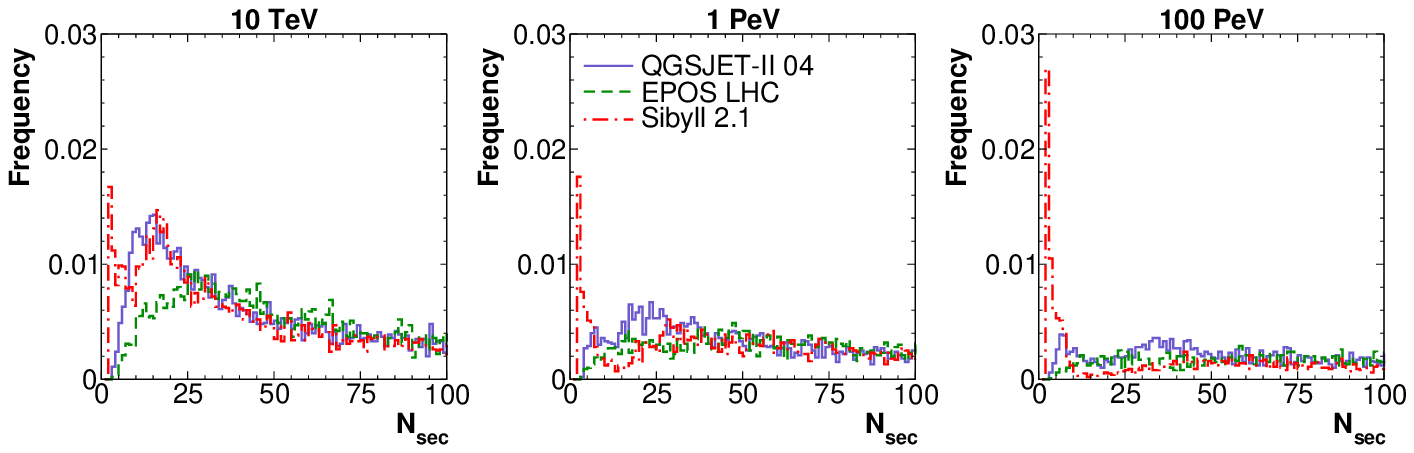}
		\caption{Comparison between the Sybill, QGSJET and EPOS LHC predictions for the number of secondaries considering $p$-Air (upper panels) and $Fe$-Air (lower panels) collisions at three different values for the primary energy.}
		\label{p_nsec_dist}
	\end{figure}

	Initially, in Fig. \ref{p_nsec_dist} (upper panels) we present the predictions for the  distributions of number of secondaries $N_{sec}$ produced in $p$-Air  collisions considering three different values for the primary energy: $E_{prim} =$ 100 GeV, 1 TeV and 100 PeV. We have that for $E_{prim} =$  100 GeV, the distinct predictions are similar. One the other hand, for $E_{prim} =$ 1 TeV, these distributions are different, with the presence of two peaks: one at small values of
$N_{sec}$, associated to diffractive events and other at larger values of $N_{sec}$, related to non - diffractive events. Finally, 	for $E_{prim} =$ 100 PeV ($\sqrt{s_{NN}} \approx 14$ TeV)  the distributions extend up to thousands of secondary particles, and the peaks  become more evident. In the case of Sibyll 2.1, the peak at low $N_{sec}$ is suppressed. In Fig. \ref{p_nsec_dist} (lower panels) we shown the $N_{sec}$ distributions for { $Fe$-Air} collisions. In this case we have that the predictions of the different models for the peak at low - $N_{sec}$ are very distinct, being enhanced in the Sibyll case and  suppressed for both QGSJET-II 04 and EPOS LHC with the increasing of the primary energy.

	\begin{figure}[t]
		\includegraphics[width=0.49\textwidth]{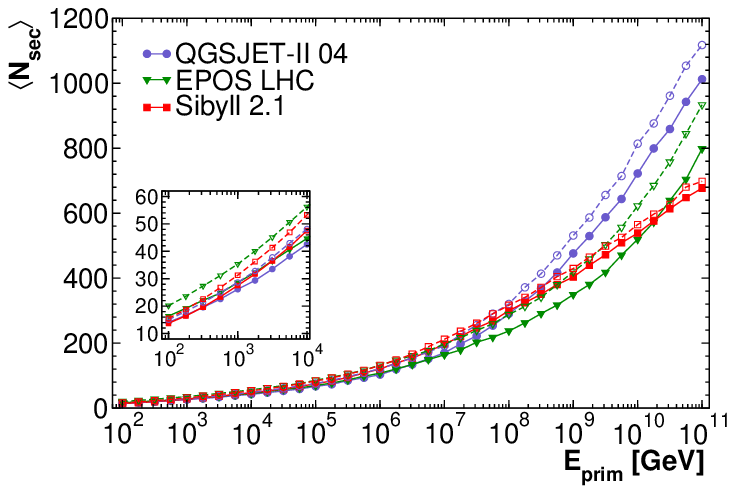}
		\includegraphics[width=0.49\textwidth]{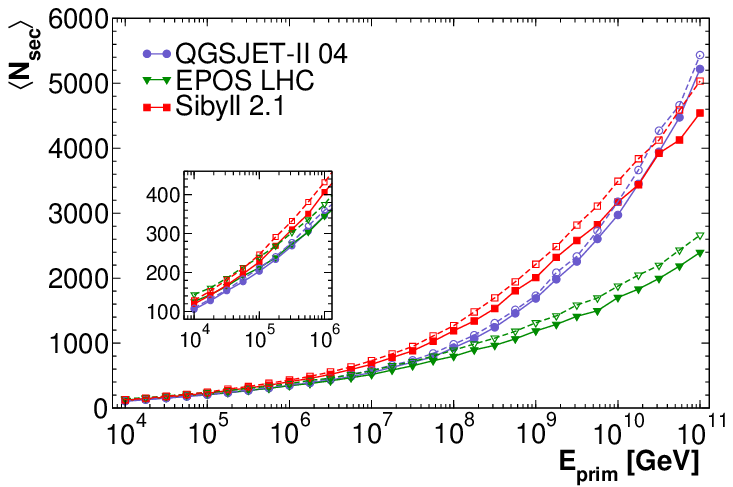}
		\caption{Predictions of the different hadronic generators for the primary energy dependence of the average number of secondaries produced in { $p$-Air} (left panel) and { $Fe$-Air} (right panel) collisions. The solid (open) symbols represent simulations including diffractive interactions (non-diffractive events only).}
		\label{nsec_media}
	\end{figure}

The predictions of the different hadronic generators for the  dependence of the average number of secondaries produced in { $p$-Air} collisions on the primary energy are presented in  Fig. \ref{nsec_media} (left panel). The full simulations, including diffractive and non - diffractive events, are  represented by solid symbols. On the other hand, the simulations including only non - diffractive events are represented by open symbols. Initially, lets compare the predictions of the different hadronic generators for the full simulations. In this case we have that for $E_{prim} \approx 10$ PeV, the  EPOS LHC and QGSJET-II 04 generators predict, on average, very similar values for the number of secondaries. As the energy increases, differences become significant, with the QGSJET-II 04 predicting the largest number of secondaries at the highest energies, while Sibyll 2.1 predicts the smaller value. One have that  the relative difference between the hadronic generators is smaller after the LHC tuning. In particular, the  QGSJET prediction for the number of secondaries produced at ultra - high energies in { $p$-Air} collisions was substantially reduced after the LHC tune. Our predictions for $Fe$ - Air collisions are presented in Fig.  \ref{nsec_media} (right panel). In this case, Sibyll 2.1 predicts  the largest number of secondaries for primary energies smaller than  10 EeV. At higher energies, the  QGSJET-II 04 predictions are  slightly larger than Sibyll 2.1, with the EPOS LHC predicting the lowest number of secondaries in the whole considered energy range. As already observed in $p$-Air collisions, the QGSJET-II 04 and EPOS LHC results are very similar at energies lower than 10 PeV and become very distinct at higher energies.
	
	Lets now analyse the impact of the diffractive events on the average number of secondaries. As can be observed in Fig. \ref{nsec_media} (left panel) by the comparison between the  solid and dashed lines, one have that for { $p$-Air} collisions the contribution of the diffractive events is small for the Sibyll 2.1. One the other hand, these events have a non - negligible impact on the QGSJET-II 04 and EPOS LHC predictions, with the contribution of diffraction increasing with the primary energy. In contrast, in  the case of the predictions for { $Fe$-Air} collisions, presented in Fig. \ref{nsec_media} (right panel), Sibyll 2.1 receives the most significant influence of diffractive interactions on the mean secondary multiplicity, while QGSJET-II 04 one is almost not influenced by diffraction.  One have that in $p$-Air and $Fe$-Air collisions, the presence of the diffractive interactions reduces the average number of secondaries produced in the collisions. It is expected, since  diffractive interactions produce less secondaries than non-diffractive collisions. Consequently, it is expected that the sample of collisions without diffraction has more secondaries on the average.

	\begin{figure}[t]
		\includegraphics[width=0.49\textwidth]{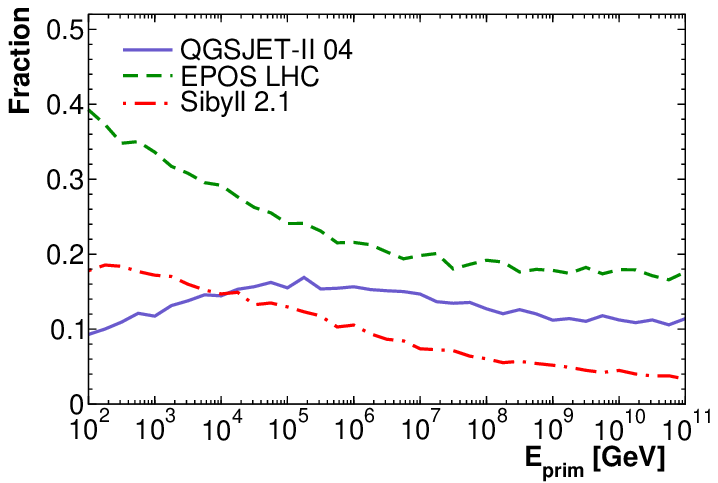}
		\includegraphics[width=0.49\textwidth]{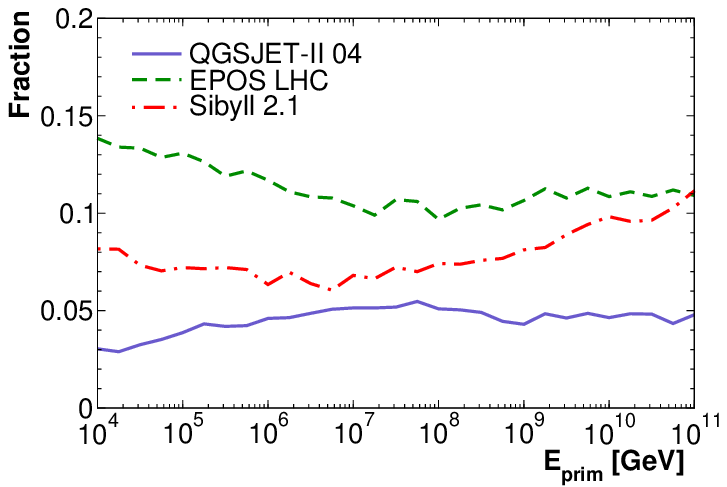}
		\caption{Fraction of diffractive events in  { $p$-Air} (left panel) and { $Fe$-Air} (right panel) collisions.}
		\label{fracdiff}
	\end{figure}

In order to estimate how the contribution of the diffractive events change with the energy, lets consider the  relative probability of diffraction, which is related to the ratio of the diffractive to the total cross sections and determines how many diffractive events are expected in a sample of collisions. The relative probabilities of diffraction predicted by the different hadronic generators for {$p$-Air} and { $Fe$-Air} collisions at different primary energies are presented in Fig. \ref{fracdiff}.  For { $p$-Air} collisions (left panel), the  EPOS LHC predicts  the largest fraction of diffractive events in the whole energy range, being $\approx$ 40\% at 100 GeV and   18\% for 100 EeV. On the other hand, Sibyll 2.1 predicts that the contribution of the diffractive events is of the order of 20\%  at 100 GeV, decreasing at larger energies. In particular,   Sibyll 2.1 predicts the smaller contribution of these events at ultra - high energies. Finally,  QGSJET-II 04 shows a different behaviour, increasing from 10\% at 100 GeV to 17\% at 100 TeV and then decreasing for 12\%  at 100 EeV. 
The results for { $Fe$-Air} collisions are presented in Fig. \ref{fracdiff} (right panel). As observed in $p$-Air collisions,  EPOS LHC predicts the largest fraction of diffractive events at all primary energies. However, the magnitude of this contribution is smaller, with values around 13\% in the energy range considered. The Sibyll 2.1 and QGSJET-II 04 predictions also are smaller in comparison to the $p$-Air collisions, with the Sibyll 2.1 one increasing with the energy and predicting similar values to the EPOS LHC at ultra - high energies.

	\begin{figure}[t]
		\includegraphics[width=0.49\textwidth]{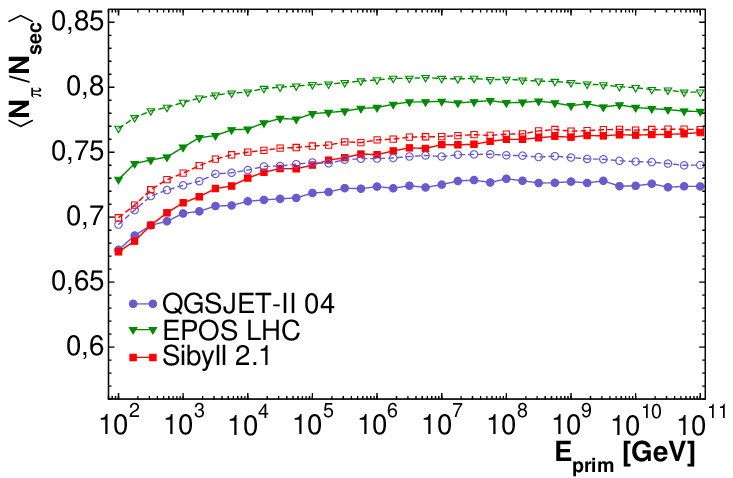}
		\includegraphics[width=0.49\textwidth]{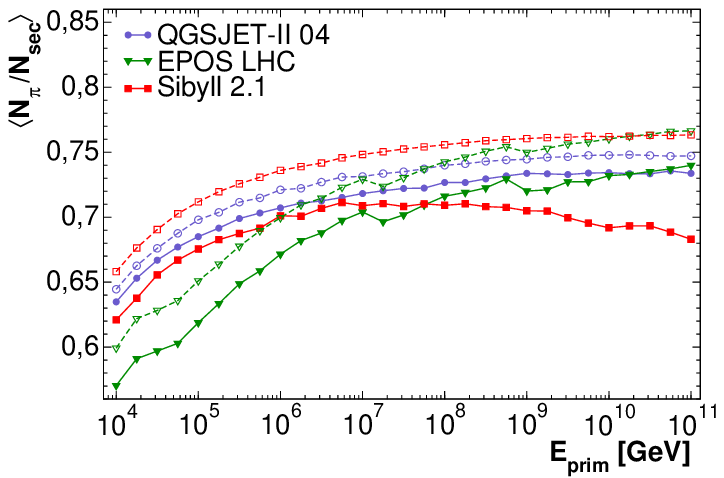}
		\caption{Energy dependence of the average fraction of pions produced in { $p$-Air} (left panel) and {$Fe$-Air} (right panel) collisions.}
		\label{fracpi}
	\end{figure}
	
	In Fig. \ref{fracpi} we present the predictions of the different hadronic generators for the energy dependence of the fraction of pions produced in {$p$-Air} (left panel) and {$Fe$-Air} (right panel) collisions, which is given by the ratio between the number of pions produced and the total number of secondaries. We have that EPOS LHC predicts the largest fraction of pions for all primary energies, with the diffractive events decreasing the fraction. On the other hand,  as expected from the Fig. \ref{fracdiff}, the fraction of pions predicted by the  Sibyll 2.1 is almost not influenced by the diffractive events. The predictions for {$Fe$-Air} collisions are presented in Fig. \ref{fracpi} (right panel). In this case, we have that the fraction of pions predicted by the three hadronic generators is smaller than in { $p$-Air} collisions and the QGSJET-II 04 predicts the largest fraction in the energy range considered. Moreover, we have that the Sibyll 2.1 prediction at large energies is strongly influenced by the diffractive events.

	\begin{figure}[t]
		\includegraphics[width=0.59\textwidth]{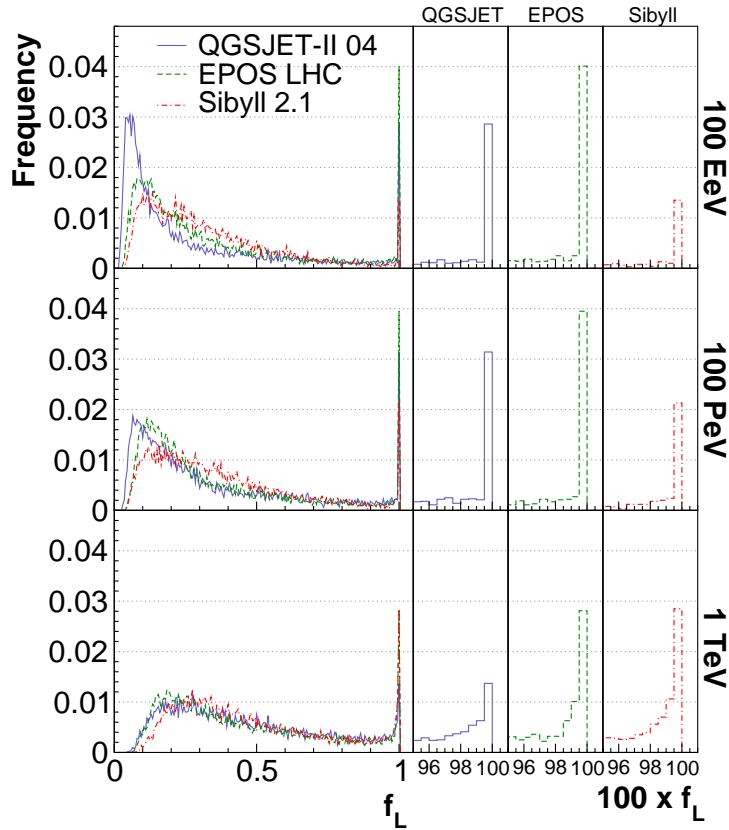}
		\caption{Predictions of the hadronic generators for the elasticity distribution considering {$p$-Air} collisions at three representative primary energies.}
		\label{p_elast_dist}
	\end{figure}

	\begin{figure}[t]
		\includegraphics[width=0.49\textwidth]{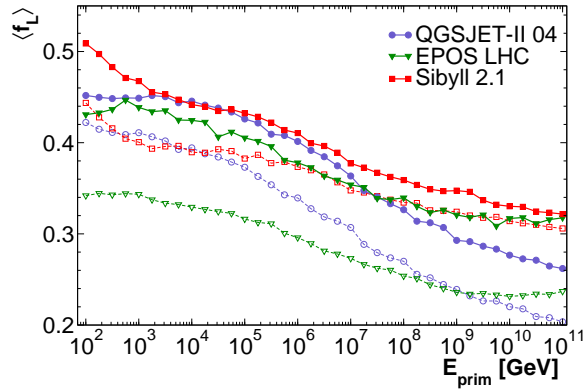}
		\caption{Energy dependence of the average elasticity considering  { $p$-Air} collisions.}
		\label{elast_media}
	\end{figure}

	In Fig. \ref{p_elast_dist} we present the predictions for the elasticity distribution considering { $p$-Air} collisions and three representative primary energies: 1 TeV, 100 PeV and 100 EeV. The large $f_L$ region is detailed in the small plots on the right. We can verify the presence of {a} peak for $f_L \approx 1$, which is related to the diffractive interactions, where the leading particle carry most of the primary energy. At small energies ($E_{prim} = 1$ TeV), the predictions of the different models are similar, except by the fact that the diffractive peak predicted by the QGSJET - II 04 is smaller in comparison to the other models. As the energy increases, the distributions starts to be different, with the QGSJET - II 04 predicting the largest fraction of non - diffractive events at small - $f_L$, which is associated to the fact that this model predicts the largest number of secondaries at high energies [See Fig. \ref{nsec_media} (left panel)]. We have that the EPOS LHC and QGSJET - II 04 predict the increasing of the peak for $f_L \approx 1$. In contrast, Sibyll 2.1 predicts that it becomes smaller at large energies, in agreement with the results presented in Fig. \ref{fracdiff}. Finally, in order to investigate the influence of the diffractive interactions in the average elasticity $\langle f_L \rangle$ {in $p$-Air collisions}, in Fig.  \ref{elast_media} we present the predictions for the energy dependence of $\langle f_L \rangle$. We have that the presence of the diffractive events implies a higher average elasticity, since these events populate the region of large  $\langle f_L \rangle$. Moreover, the largest impact is in the EPOS LHC predictions, which is associated to the fact that this model predicts the largest peak for $f_L \approx 1$.
	
Our results for $p$-Air and $Fe$-Air collisions demonstrated that the diffractive interactions modify the magnitude of the number of secondaries and the inelasticy of these collisions. Moreover, our results indicated that the distinct treatment of these interactions, present in the different hadronic generators, implies a non - negligible theoretical uncertainty in the predictions. In the next Section, we will expand our analysis for air shower observables.

% ==================================================================== %
\section{Impact of the diffractive interactions on Shower Observables}
\label{shower}

	In order to investigate the impact of the  diffractive interactions on the air shower observables we will use the software CORSIKA (Version 7.4.005) to simulate air showers generated by primary protons and iron nuclei with energies of $10^{17}$ and $10^{20}$ eV, reaching the atmosphere with a zenith angle of 60º. The simulations have been performed considering  two distinct configurations: (a) full simulations, which include diffractive and non - diffractive interactions  in the shower development, and (b) non  - diffractive (ND) simulations generated removing the diffractive interactions of the shower development. For the description of the hadronic interactions we will consider the same models discussed in the previous Section.

	\begin{figure}[t]
		\includegraphics[width=\textwidth]{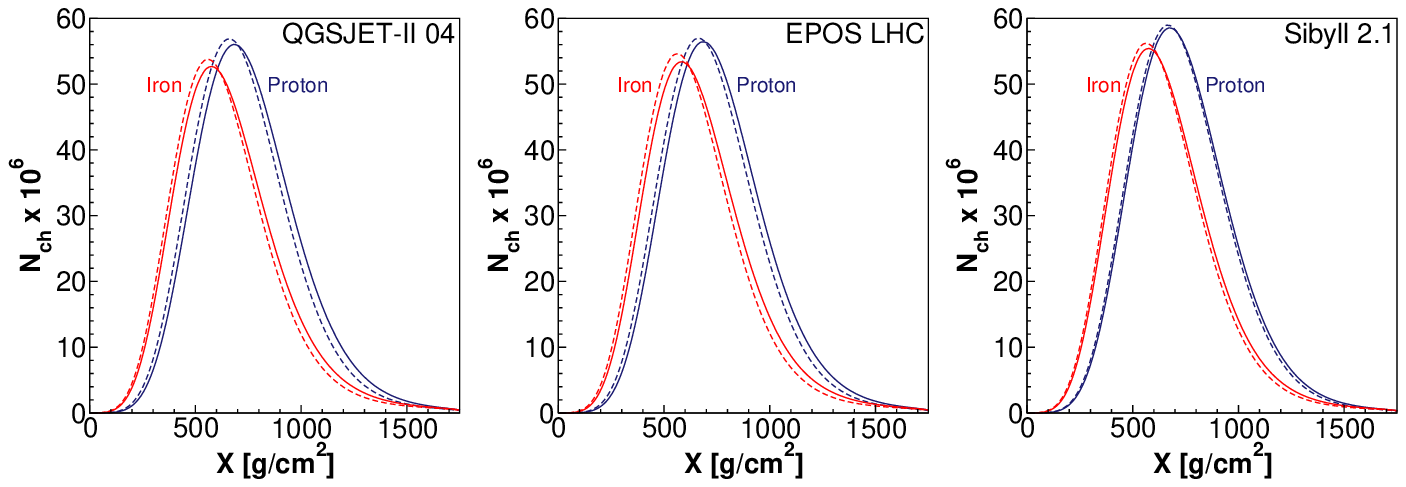} \\
         \includegraphics[width=\textwidth]{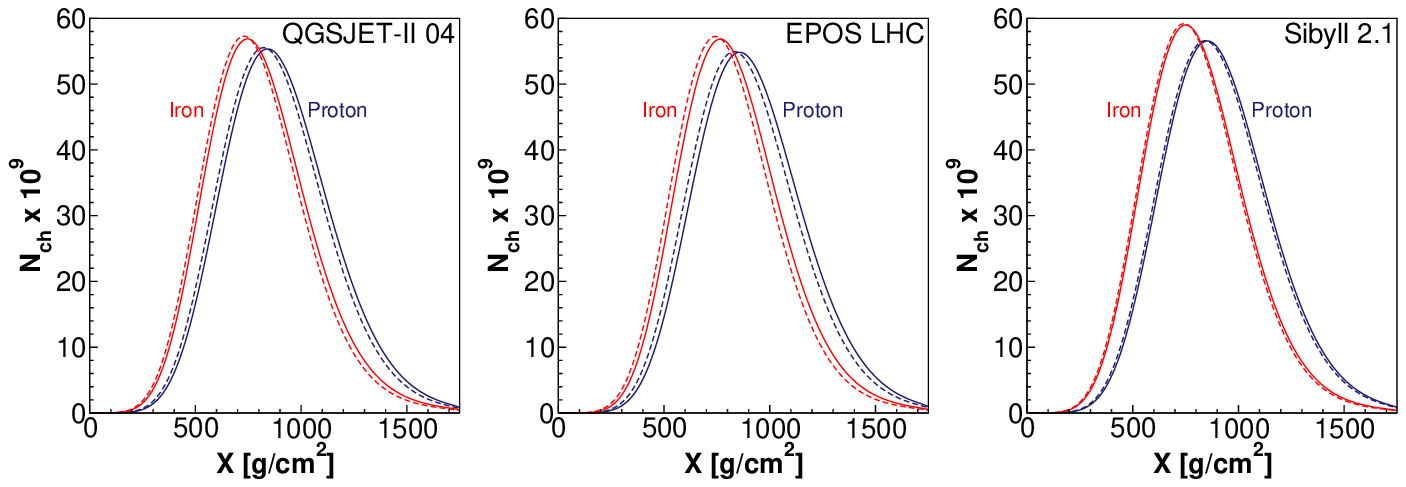}
		\caption{Average longitudinal profiles of charged particles for showers generated by protons and iron nuclei with primary energies of $10^{17}$ eV (upper panels) and $10^{20}$ eV (lower panels). Solid (dashed) lines represent the full (non - diffractive) simulations.}
		\label{long_ch_17}
	\end{figure}

In Fig. \ref{long_ch_17} we present the predictions for the mean longitudinal profiles of charged particles for showers generated by protons and iron nuclei at $10^{17}$ eV (upper panels) and $10^{20}$ eV (lower panels). The solid lines represent showers generated including the non - diffractive and diffractive interactions (full simulations), while the dashed lines represent showers where the diffractive interactions were removed of the shower development (non - diffractive simulations). The impact of the diffractive interactions is very clear: the presence of the diffractive events implies that the average number of particles is smaller  and the maximum is shifted to higher atmospheric depths,  i.e. the air showers develop slower in the atmosphere, in agreement with the results obtained in Ref. \cite{garcia1}. Moreover, we have that the Sibyll 2.1 predictions are almost insensitive to the inclusion of the diffractive interactions, which agrees with the fact that the contribution of these interactions at high energies is small in this model.

	\begin{table}[t]
    \centering
		\begin{tabular}{||l| c | c | c | c||}
		\hline
		\hline
	$E_{prim}$ & Primary & \: QGSJET-II 04 \: & \: EPOS LHC \: & \: Sibyll 2.1 \: \\
			\hline \hline
	{	$10^{17}$ eV} &	 \: p \:  & 24.5 $g/cm^2$ & 23.3 $g/cm^2$ & 8.3 $g/cm^2$ \\
				& \: Fe \: & 17.0 $g/cm^2$ & 20.0 $g/cm^2$ & 11.2 $g/cm^2$ \\
				\hline
				\hline
	{	$10^{20}$ eV} &	 \: p \:  & 19.6 $g/cm^2$ & 26.0 $g/cm^2$ & 10.9 $g/cm^2$ \\
				& \: Fe \: & 16.5 $g/cm^2$ & 22.5 $g/cm^2$ & 8.3 $g/cm^2$ \\
				\hline
				\hline
		\end{tabular}
		\caption{Predictions for the shift in $\langle X_{max} \rangle$   due to the presence of the diffractive events.}
    \label{xmax_shift_17}
	\end{table}
	
	\begin{figure}[t]
		\includegraphics[width=0.3\textwidth]{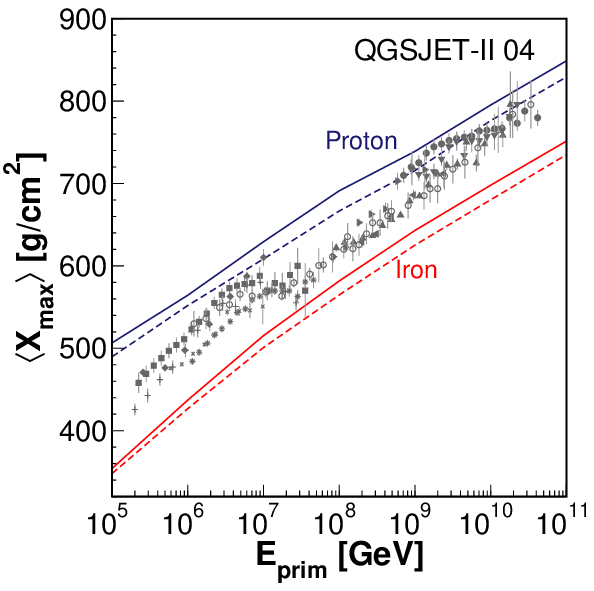}
		\includegraphics[width=0.3\textwidth]{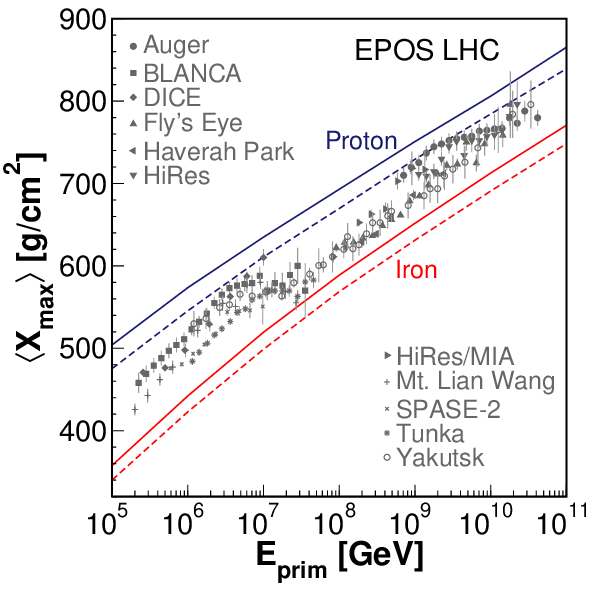}
		\includegraphics[width=0.3\textwidth]{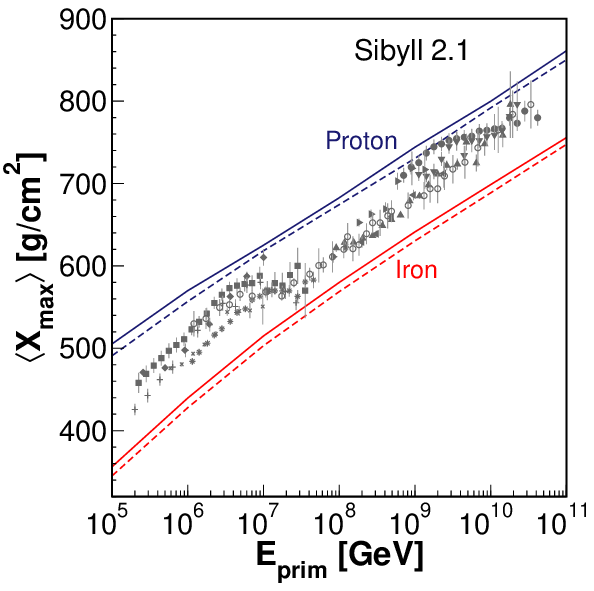}
		\caption{Depth of maximum $\langle X_{max} \rangle$ as a function of the primary cosmic ray energy. Solid (dashed) are from simulations including (removing) the diffractive interactions of the air shower development.  }
		\label{xmax}
	\end{figure}

The influence of the diffractive interactions also can be estimated by the analysis of the shift in the depth of maximum $\langle X_{max} \rangle$, given by $\langle X_{max}^{(full)} \rangle - \langle X_{max}^{(ND)} \rangle$. In Table \ref{xmax_shift_17} we present our results for  this quantity.	We can see that the influence of the diffractive interactions on the profiles is smaller in the case of Sibyll 2.1, for both primaries. This is related to the fact that Sibyll 2.1 usually produces less diffractive events (See Fig. \ref{fracdiff}) in hadron-nucleus collisions than the other models. Additionally, we have that the showers generated using the EPOS LHC are more influenced by diffraction than those generated by the other models. This also can be verified in Fig. \ref{xmax}, where the energy dependence of $\langle X_{max} \rangle$ is presented. Moreover, we have that the impact of the diffractive interactions on this observable  is almost  energy-independent for the three considered models of hadronic interactions.
We have that the impact of the diffractive interactions in the shower maximum position depends of the treatment of the diffractive physics and it is of the order of the typical experimental precision for the  $\langle X_{max} \rangle$ measurements. Such uncertainty has as consequence the degradation of the accuracy of other quantities, as for example, the composition of the UHECR \cite{osta}.

	\begin{figure}[t]
		\includegraphics[width=\textwidth]{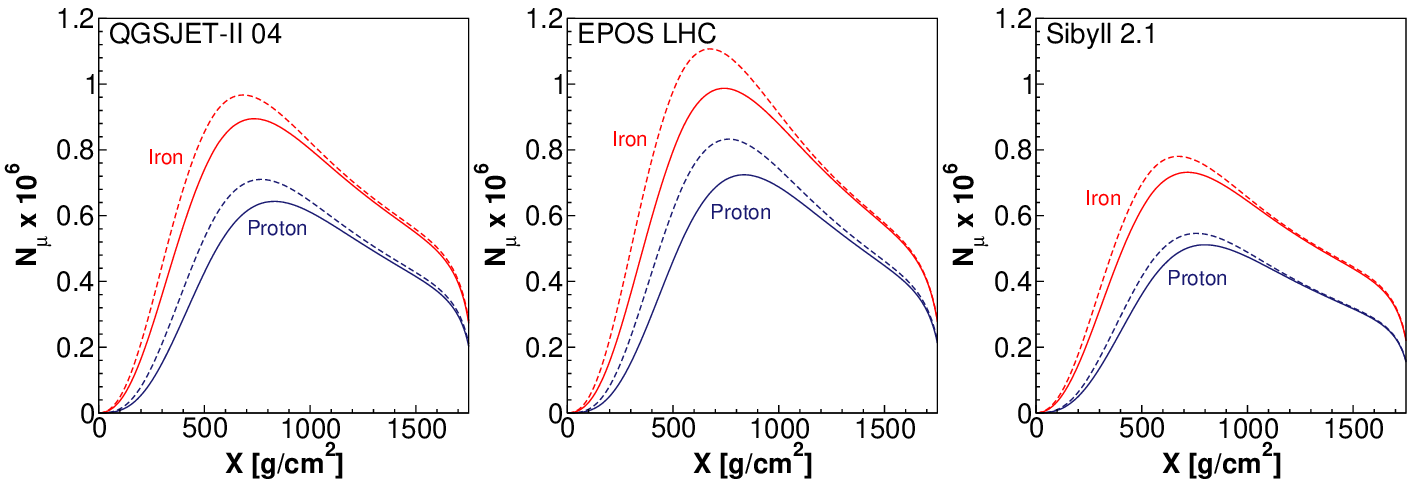} \\
		\includegraphics[width=\textwidth]{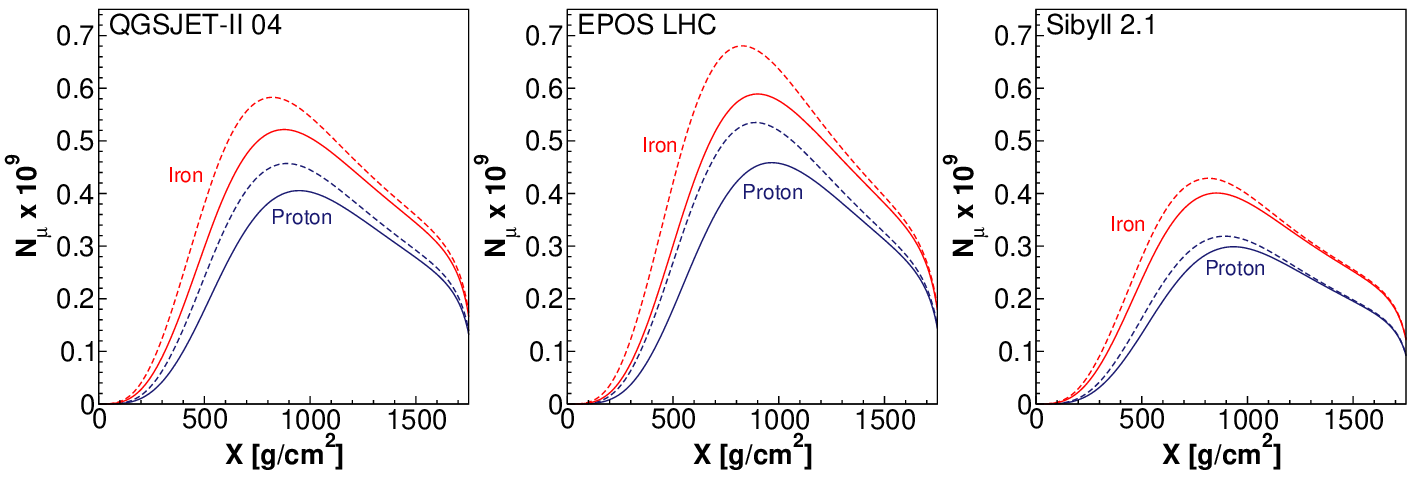}
		\caption{Predictions for the average longitudinal profiles of muons for air showers generated by protons and iron nuclei with primary energies of $10^{17}$ eV (upper panels) and  $10^{20}$ eV (lower panels) . Solid (dashed) lines represent the full (non - diffractive) simulations.}
		\label{long_mu_17}
	\end{figure}
	
Lets now analyse the impact on the muonic content of the air showers, which is known to be related to the primary composition and also to the properties of the hadronic interactions on the shower development. In particular,  Auger collaboration has recently measured the depth of maximum production of muons $X_{max}^{\mu}$ in high energy air showers and has verified that QGSJET-II 04 bracketed the data with simulations of proton and iron induced showers, while EPOS LHC underestimates values of $X_{max}^{\mu}$ \cite{Aab:2014dua}. Auger collaboration also showed that both models underestimate the muon content that reach the ground \cite{Aab:2014pza}, with the discrepancies being larger for the QGSJET-II 04 predictions than those from the EPOS LHC model.
In Fig. \ref{long_mu_17}  we present the predictions of the different hadronic generators for the longitudinal profiles of the muonic component for air showers generated at $10^{17}$ eV (upper panels) and $10^{20}$ eV (lower panels). Qualitatively, the influence of diffraction is the same observed  for the profiles of charged particles: simulations without diffractive processes produce more muons and the profiles are shifted towards lower atmospheric depths. It is worth note that EPOS LHC predicts more muons than the other models and that its predictions for the profiles are more dependent on the diffractive interactions.

	\begin{figure}[t]
		\includegraphics[width=\textwidth]{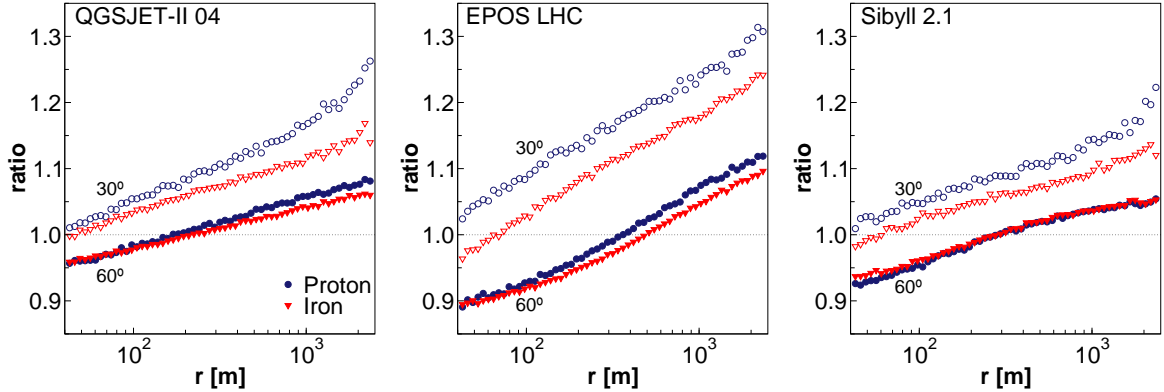}
		\caption{Ratio  between the predictions for muon densities obtained considering only non - diffractive interactions and those derived including diffractive and non - diffractive interactions as a function of the distance to the shower core. Showers initiated by protons (iron nuclei) are represented by blue circles  (red triangles). Solid (open) symbols denote showers initiated by primary reaching the atmosphere with a zenithal angle of 60º (30º). }
		\label{late_mu_17}
	\end{figure}

Finally, lets analyse the influence of the diffractive interactions on the  density of muons that hit the ground ($\rho_{\mu}$).  In particular, we  estimate the ratio  between the predictions for $\rho_{\mu}$ obtained considering only non - diffractive interactions and those derived including diffractive and non - diffractive interactions. In  Fig.  \ref{late_mu_17} we present our results for the ratio $\rho_{\mu}^{(ND)}/\rho_{\mu}^{(full)}$ as a function of distance to the core shower, assuming a primary energy of $10^{17}$ eV and an observation level of 1400 m (Pierre Auger Observatory), which corresponds to an atmospheric depth of 1760 $g/cm^2$ { for a zenith angle of 60º. For showers generated with a zenith angle of 30º degrees, such altitude corresponds to an atmospheric depth of $\approx$ 1000 $g/cm^2$.}
We present results assuming  { both values} for the zenith angle. We have that the ratio increases with distance to the shower core and that the influence of the diffractive interactions becomes non - negligible at large distances, specially for small values of the zenith angle.

A final comment is in order. After the completion of our study, a new version of the event generator Sibyll have been released \cite{sibyll23}. One of the implications of the modifications implemented in Sibyll 2.3 is the enhancement of the diffractive interactions. As a consequence, the new version predicts, on average, less secondaries in the whole energy range considered for both primaries. One have verified that our results are not strongly modified. The main modification is that the differences between non - diffractive and full simulations of the distinct observables  are enhanced by a factor smaller than 1.25 ($\lesssim$ 25\%) in comparison with the former version Sibyll 2.1, for all energies and primaries. Therefore, our main conclusions about the impact of the diffractive interactions remain valid using the Sibyll 2.3.

% ==================================================================== %
\section{Summary}
\label{conc}

In this paper we invertigated the impact of the diffractive interactions on distinct observables for $p$-Air and $Fe$-Air collisions as well as in ultra - high energy cosmic ray interactions. Our results demonstrated that the distinct phenomenological models, present in the CORSIKA package, predict a different magnitude for the fraction of diffractive events and for its energy dependence. As a consequence, the influence of these interactions on the number of secondaries and fraction of pions is strongly model dependent. We demonstrated that the predictions of these models for the elasticiy are distinct, which directly modifies the air shower development.
Our results for $\langle X_{max} \rangle$ indicated  that the impact of the diffractive interactions on this observable  is almost  energy-independent for the three considered models of hadronic interactions and it is of the order of the typical experimental precision for the  $\langle X_{max} \rangle$ measurements. Moreover, we shown that the average longitudinal profile of muons is sensitive to the diffractive interactions, in particular for small zenithal angles. Our results indicated that the diffractive  interactions has a non - negligible influence on the observables and that the treatment of the diffractive physics is an important source of uncertainty in the description of the extensive air showers.

% ==================================================================== %
\begin{acknowledgments}

	This work was  financed by the Brazilian funding agencies CAPES, CNPq and FAPERGS.
	
\end{acknowledgments}

% ==================================================================== %

\end{document}